\documentclass[twocolumn,prl,aps]{revtex4}

\usepackage{epsf}

\begin{document}

\title{On a liquid drop  "falling" in a heavier miscible fluid}

\author{Paul K. Buah-Bassua$\text h^3$, Ren\'e Roja$\text s^1$, Stefania Residor$\text i^1$, F.T. Arecch$\text i^2$}
\address{$^1$ Institut Non-Lin\'eaire de Nice, 1361 Route des Lucioles, 06560 Valbonne, France
\\
$^2$Phys. Dept., University of Florence and
Istituto Nazionale di Ottica Applicata, Largo E. Fermi 6, 50125 Florence, Italy
\\
$^3$ Physics Dept. of the University of Cape Coast, Cape Coast, Ghana}

\date{\today}

\begin{abstract}
We report a new type of drop instability, where the density difference between the drop and the solvent is negative. We show that the drop falls inside the solvent down to a minimum height, then fragmentation takes place and secondary droplets rise up to the surface. We have developed a theoretical model that captures the essential of the phenomenon and predicts the correct scalings for the rise-up time and the minimum height.   
\end{abstract}

\maketitle

\smallskip

Pacs: 47.20.-k,
47.20.Bp,
47.32.Cc,Ê 
68.05.-nÊ 

\bigskip

In the recent years, the physics of liquid drops has been faced with fundamental questions that are becoming more and more relevant to practical agricultural and industrial applications \cite{eggers}. In particular, it seems such fundamental relevance to the dynamics of the drops and the interfaces are ruled by an interfacial surface tension that would be related to the transient presence of velocity gradients between the liquid of the drop and the one composing the solvent \cite{korteweg}. Recently, we have demonstrated that when a drop of liquid is deposited over the surface of the same liquid, it falls down inside the solvent because the energy associated to its surface tension against air is instantaneously converted into kinetic energy. As a consequence, a very fast fluid injection takes place as the drop touches the surface of the solvent \cite{anilkumar,residori}. Universal scaling laws apply, relating the initial velocity transferred at the injected drop to the minimum height at which it stops inside the solvent \cite{residori}.

\begin{figure} [h!]
\centerline{\epsfxsize=8 truecm \epsffile{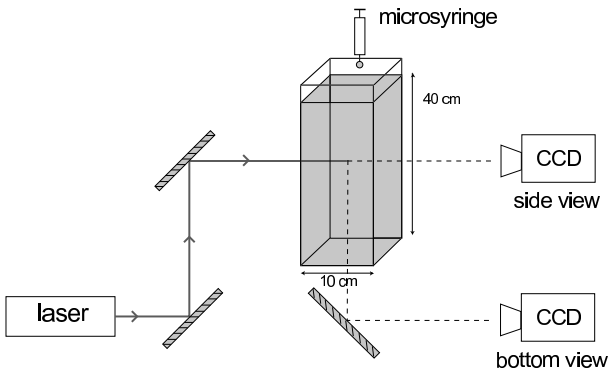}}
\caption{Experimental setup: a solid state laser beam ($\lambda=532$ $nm$) is shone laterally onto the cell; fluorescence from the drop is recorded by a CCD camera. 
\label{setup}}
\end{figure}

It is a well-known phenomenon that, when a liquid drop falls inside a miscible fluid and the density difference between the drop and the solvent $\Delta \rho$, is slightly positive (the drop is heavier than the solvent), then the drop fragments into smaller and smaller droplets \cite{thomson}. At longer times, the process is washed out by the diffusion of the drop liquid into the solvent. Recently, we have shown that this hydrodynamical instability is ruled by two non dimensional numbers, the fragmentation number F and the Schmidt number S \cite{arecchi1,arecchi2} and displays fractal properties in the statistics of the drop fragments \cite{arecchi3}.

Here, for the first time we report a set of experiments performed in the case of negative $\Delta \rho$ (the drop is lighter than the solvent) and we show that, despite the negative sign of the density difference, the drop does indeed "fall" inside the solvent, at least initially, when a fast injection takes place because of the almost instantaneous conversion of surface energy into kinetic impulsion.
When going inwards the solvent, the drop develops a vortex ring, which expands and falls down until the initial impulsion is dissipated by viscosity.
When the ring stops, a new instability takes place that leads to the fragmentation of the ring into smaller droplets. This is a Rayleigh-Taylor instability due to density difference between the drop and the solvent \cite{tritton}. When $\Delta \rho>0$ the drop fragments continue their descent
down inside the solvent, but for $\Delta \rho<0$, the density difference being negative, an inversion of velocity takes place and the drop fragments rise up towards the surface of the solvent. Thus, the instability is equivalent to a Rayleigh-Taylor instability but with the sign of gravity reversed.

\begin{figure} [h!]
\centerline{\epsfxsize=7 truecm \epsffile{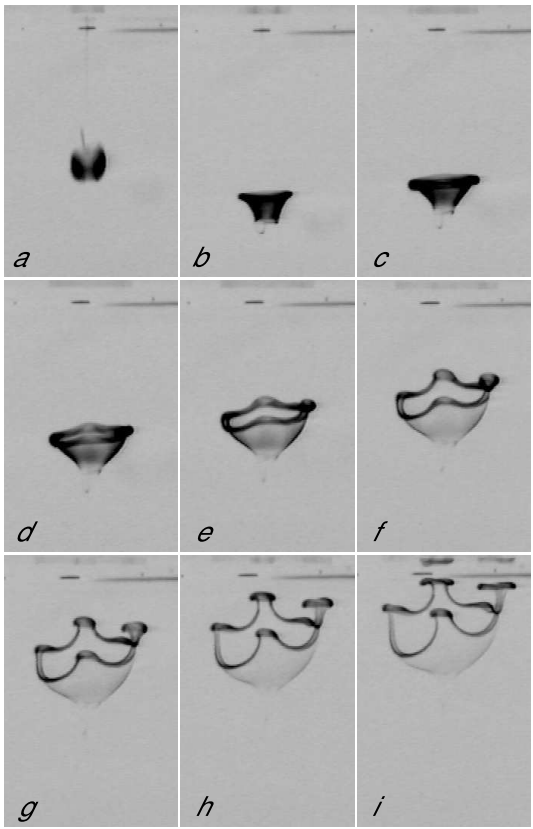}}
\caption{Frame assembly, showing the drop injection, the ring formation and the fragment rise-up; a) $t=0.08$ $s$, b)  $t=0.20$ $s$, c) $t=0.32$ $s$, d) $t=0.44$ $s$, e) $t=0.56$ $s$, 
f) $t=0.68$ $s$, g) $t=0.80$ $s$, h) $t=0.92$ $s$, i) $t=1.04$ $s$.
\label{assembly}}
\end{figure}

The experimental setup, shown in Fig.\ref{setup}, consists of a glass cell with a base of $10$x$10$ $cm^2$ and $40$ $cm$ high, mounted on a rigid metallic support. We have studied the behavior of different fluids, namely the solvent was made up of distilled and purified water doped at $10$, $15$, $25$ $\%$ Glycerol and the drop was made up of distilled and purified water with a Glycerol concentration varying in between $0$ and $25$ $\%$. The drop, which is formed at the needle of a high precision Hamilton microsyringe, has a volume $V$ that can be adjusted from $1$ to $10$ $\mu l$, with an accuracy of a few percents. Once formed, the drop is deposited adiabatically, by means of a micrometric translation stage, on the free surface of the solvent. Side and bottom views of the drop inside the solvent are recorded by means of a CCD camera and a solid-state laser illumination ($\lambda=532$ $nm$). The drop is slightly doped with Fluorescein at the purposes of visualization.

\begin{figure} [h!]
\centerline{\epsfxsize=8 truecm \epsffile{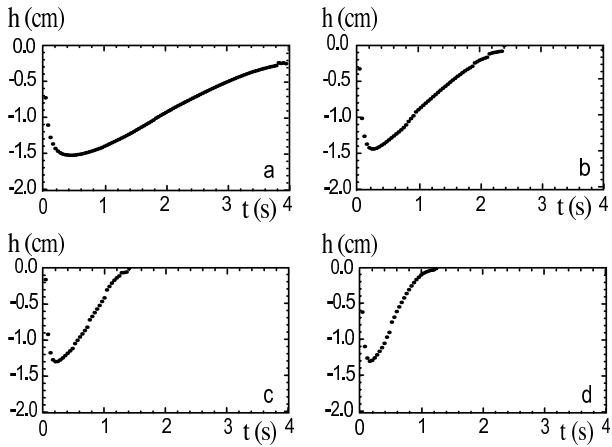}}
\caption{Drop height $h$ as a function of time for $V=4$ $\mu l$;  $\Delta \rho=$ a) $0.01325$,
b) $0.0265$, c) $0.03975$ and d) $0.04505$ $g/cm^3$. 
\label{h-t-vol}}
\end{figure}

\begin{figure} [h!]
\centerline{\epsfxsize=8 truecm \epsffile{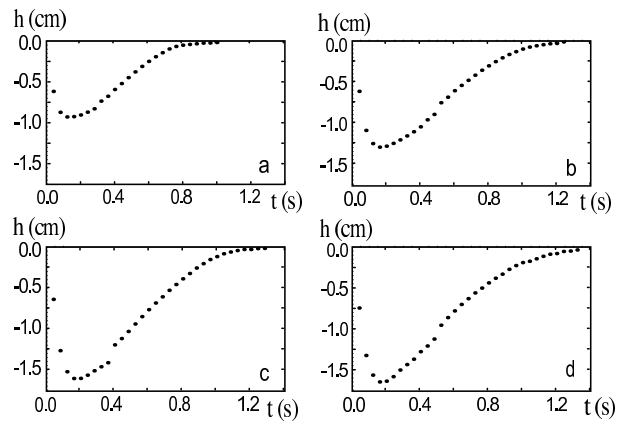}}
\caption{Drop height $h$ as a function of time for $\Delta \rho=0.04505$ $g/cm^3$;
$V=$ a) $2$, b) $4$, c) $6$ and d) $8$ $\mu l$. 
\label{h-t-rho}}
\end{figure}

A typical behavior observed for a $V=2$ $\mu l$ drop doped at $15$ $\%$ Glycerol and falling in a $25$ $\%$ Glycerol 
doped solvent ($\Delta \rho=0.053$ $g/cm^3$ ) is shown in the assembly of Fig.\ref{assembly}. The label in each frame 
corresponds to the time sequencing, where the camera acquisition rate is of $25$ frames/$sec$. We can distinguish the 
fast injection of the drop, the ring formation, its undulation and the subsequent fragmentation into four droplets, 
then rising-up towards the free surface of the solvent. It is worth to note also that, when the ring expands, it remains 
attached to a convex membrane. In the case of positive $\Delta \rho$ a similar phenomenon was also observed and called Òturban instabilityÓ \cite{arecchi1}. In that case the curvature of the membrane was in the opposite direction with respect to the present case. Note that the turban instability has been observed also in the case of immiscible fluids
\cite{boratav}.

We have performed several experiments by changing the drop volume $V$ and the density difference $\Delta \rho$. For each 
set of experiments we have recorded several movies following the drop evolution and for each recorded movie we have 
performed the following processing.
We have binarized all the frames by choosing a unique threshold intensity and by checking that this one minimizes the 
discontinuities between each frame and its successive.
Then, on each frame we identify the center of mass of the drop, we record its coordinates
and we follow its trajectory until the drop stops its descent and starts to rise up breaking 
into fragments. At this point, we choose only one fragment and follow its motion by recording the coordinates of its 
center of mass. The evolution of the longitudinal coordinate,
$h$, of the center of mass is plotted as a function of time for a fixed drop volume, $V=4$ $\mu l$, and for different 
$\Delta \rho$ (Fig.\ref{h-t-vol}) and for a fixed 
$\Delta \rho=0.04505$ $g/cm^3$ and different drop volumes, $V=2,4,6,8$ $\mu l$ (Fig.\ref{h-t-rho}).

From Fig.\ref{h-t-vol} and \ref{h-t-rho}, we can see that, once the drop has evolved into a vortex ring, it stops at a minimum height, $h_{min}$, which is mainly ruled by the initial drop volume, $V$. On the other hand, when fragmentation takes place, the rise-up time for the secondary droplets mainly depends on the density difference, $\Delta \rho$, eventually going to infinity
for $\Delta \rho=0$. At small  $\Delta \rho$ the rise-up time is very long, while it shortens as $\Delta \rho$ increases. The drop injection takes place even in the absence of density difference because, the two fluids being miscible, there is an "instantaneous" conversion of
the energy associated to surface tension into kinetic energy \cite{anilkumar,residori}.
Then, viscous dissipation slows down the motion of the drop, which asymptotically reaches the minimum height $h_{min}$.

We can describe the dynamical behavior of the drop by writing a simple model that takes into accounts buoyancy and viscous dissipation. The
equation of motion reads 

\begin{equation}
{dv \over dt} = {g \Delta \rho \over \rho} - \gamma {\nu \over r^2} v,
\label{eq1}
\end{equation}
with  $r=\kappa \sqrt [3] V$ and $\gamma$, $\kappa$ geometrical factors ($\gamma=9/2$ and $\kappa=0.62$
for a sphere \cite{tritton}).
The initial condition is given by the injection of the drop, $v(t=0)=v_0$, where
$v_0$ comes from the conversion of the drop surface tension into kinetic translational and rotational energy
\begin{equation}
{1 \over 2}m v_0^2+ {1 \over 2}I \omega^2= 4 \pi \sigma r^2,
\label{eq2}
\end{equation}
with $I=\alpha m r^2$ the inertial momentum of the drop and $\omega=\beta v_0 /r$ its frequency of rotation. If all the rotation is converted into translation, i.e., there is no
sliding, then $\beta=1$, otherwise $\beta>1$. We obtain for the initial velocity of the drop

\begin{equation}
v_0=- \sqrt{{6 \sigma \over (1+\alpha \beta^2) \rho r}}.
\label{eq3}
\end{equation}
By defining the viscous time, $\tau_\nu=r^2 / \gamma \nu$, we derive from Eq.\ref{eq1} the drop asymptotic velocity, $v_\infty$, corresponding to $dv/dt=0$,
\begin{equation}
v_\infty={\Delta \rho \over \rho}g\tau_\nu.
\label{eq4}
\end{equation} 
Integrating Eq.\ref{eq1} from $v=v_0$ to $v=0$ we obtain the drop fall-down time, $\tau_{d}$, which is the time taken by the drop to stop 
\begin{equation}
\tau_d=\tau_\nu \ln(1-{v_0 \over v_\infty}),
\label{eq5}
\end{equation} 
and the minimum height, $h_{min}$, reached by the drop before rising-up
\begin{equation}
h_{min}=v_\infty \tau_d + v_0 \tau_\nu.
\label{eq6}
\end{equation} 

As for the fragment rise-up time $\tau_{u}$, if $t  \gg \tau_\nu$ it is simply given by
\begin{equation}
\tau_{u}=- {h_{min} \over v_\infty}=\mid {v_0 \over v_\infty} \mid \tau_\nu - \tau_d, 
\label{eq7}
\end{equation} 
so that the total elapsed time is $\tau_T= \mid {v_0 / v_\infty} \mid  \tau_\nu$. 
However, as we can see from Fig.\ref{assembly}, the rising-up droplets are fragments of the initial drop, so that the asymptotic velocity to be used here has the same expression as before, Eq.\ref{eq4}, but with a volume $V / n$  that is a fraction of the initial one, where $n$ is the number of fragments.
If we take into account this correction, we have that 
\begin{equation}
\tau_{u}= (\tau_T - \tau_d)n^{2 / 3}.
\label{eq8}
\end{equation} 

\begin{figure} [h!]
\centerline{\epsfxsize=6 truecm \epsffile{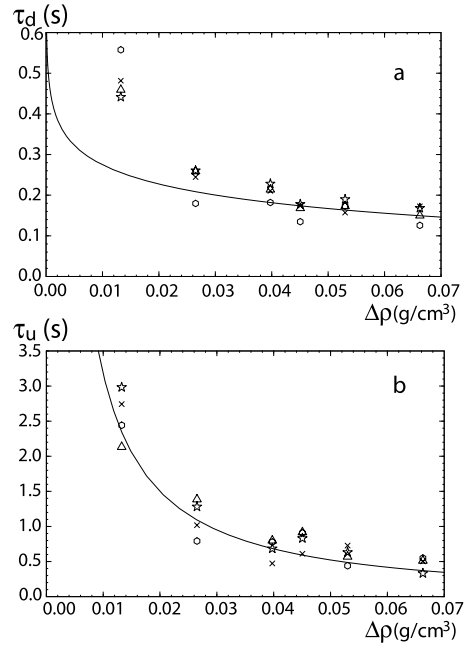}}
\caption{a) Drop fall-down time $\tau_{d}$  and b) rise-up time $\tau_{u}$ as a function of $\Delta \rho$; $V=2$ $\mu l$ circles, $V=4$ $\mu l$ triangles, $V=6$ $\mu l$ stars, $V=8$ $\mu l$ crosses. Lines are the theoretical curves for $V=5$ $\mu l$.
\label{times}}
\end{figure}

We show in Fig.\ref{times}a and Fig.\ref{times}b,  respectively,
the drop fall-down time $\tau_{d}$, and rise-up time $\tau_{u}$, as a function of $\Delta \rho$. 
From now on, we fix the parameters of the model to $\alpha \beta^2 =4$, $\gamma=6.67$ and
$\kappa=0.56$. 
We plot in Fig.\ref{times}a the theoretical prediction for $\tau_{d}$, as in Eq.\ref{eq5}. This curve fits quite well the data for $\Delta \rho>0.02$ $g/cm^3$ but presents large deviations for lower values of $\Delta \rho$. Indeed, when $\Delta \rho \rightarrow 0$ the logarithmic divergence 
does not take into account the dissipation due to the increasing radius of the vortex ring. To include such an effect a more refined model should be developed in order to describe the dynamics of the ring formation.
 As for the rise-up time $\tau_{u}$, we have plotted in Fig.\ref{times}b the experimental data by normalizing each value at $n^{2/3}$, where $n$ is the number of secondary droplets after the fragmentation has taken place. By using the expression in Eq.\ref{eq7}, we obtain a good fit of all the data. In Fig.\ref{times}b we report the curve for $V=5$ $\mu l$, the curves for the other volumes being close to this one.

\begin{figure} [h!]
\centerline{\epsfxsize=7 truecm \epsffile{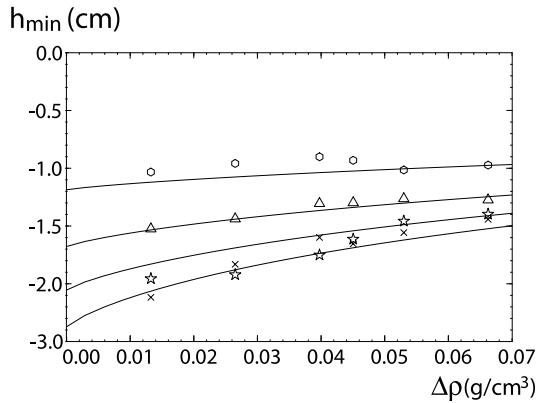}}
\caption{Minimum drop height $h_{min}$ as a function of $\Delta \rho$;
$V=2$ $\mu l$ circles, $V=4$ $\mu l$ triangles, $V=6$ $\mu l$ stars, $V=8$ $\mu l$ crosses.
\label{hmin}}
\end{figure}

The minimum height $h_{min}$ reached by the drop before rising-up is plotted in Fig.\ref{hmin} as a function of $\Delta \rho$, together with the theoretical curves, 
Eq.\ref{eq6}, for $V=2,4,6,8$ $\mu l$. We can see that the
theoretical curves are in good agreement with the experimental data.
Note that, in the limit of the experimental error, for $\Delta \rho=0$ we obtain the $h \propto V^{1 \over 2}$ scaling, in agreement with the previously reported law \cite{residori}.

\begin{figure} [h!]
\centerline{\epsfxsize=7 truecm \epsffile{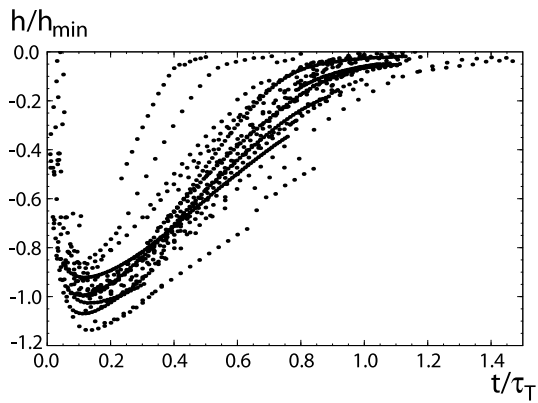}}
\caption{Reduced $h-t$ profiles for the all the experiments in
$25 \%$ Gly doped solvent.
\label{tous}}
\end{figure}

Finally, we rescale all the $h-t$ data by $h_{min}$ and $\tau_T$, and we plot the reduced profiles in Fig.\ref{tous}. 
We can see that all the drops approximately follow the same evolution law. 
The early stages of the drop injection are very similar to those observed at $\Delta \rho=0$: the drop falls very fast inside the solvent and develops a ring. Then, the ring stops because of dissipation of the initial impulsion.
At this point, the drop has reached the minimum height $h_{min}$, where a velocity reversal occurs
and where a new instability takes place leading to the fragmentation of the ring into smaller droplets. 
Being $\Delta \rho<0$, the secondary droplets rise-up towards the surface
because of buoyancy. This dynamical regime corresponds to the linear portions of the $h-t$ profiles just after $h_{min}$.
Viscous dissipation slows down the motion of the fragments, but in this dynamical regime buoyancy is dominant.

At later times, when fragments approach the surface, we observe deviations from the linear dependence of $h$ {\it vs} $t$. 
Indeed, droplets feel the presence of the boundary and behave as secondary vortex rings, each one colliding over a
wall with a longitudinal velocity component \cite{book}.
To describe the interaction of the vortex ring with the wall we can replace the wall by the specular image of the vortex ring, this one having opposite circulation
with respect to the incoming ring. Because of this interaction, the vortex ring expands and slows down, until, at later times, diffusion takes over the whole process.

In conclusion, we have reported a new type of drop instability, where secondary droplets rise up to the surface because of the negative density difference between the drop and the solvent. We have developed a theoretical model that takes into account
the initial conversion of surface tension into kinetic energy, and we show that buoyancy and viscous dissipation rule the dynamics. Even though simplified, the model captures the essential of the phenomenon and predicts the correct scalings for
the rise-up time and the minimum height reached by the drop inside the solvent.

{\it Acknowledgements}
P.K. Buah-Bassuah gratefully acknowledges the International Center for Theoretical Physics of Trieste, 
Italy, under the TRIL program, for partially supporting his visit at INLN and CNRS France grant. R. Rojas acknowledges financial support from Beca Presidente
de la Rep\'ublica of the Chilean Government.

\end{document}